\documentstyle[aps]{revtex}
\begin{document}
\title{
Elimination of unoccupied state summations\\ in {\it ab initio} self-energy calculations for large supercells
   }
\author{Lucia Reining}
\address{  Laboratoire des Solides Irradi\'es,
   URA 1380 CNRS - CEA/DTA/DECM
   Ecole Polytechnique, F-91128 Palaiseau, France }

\author{Giovanni Onida}
\address{  Laboratoire des Solides Irradi\'es,
   URA 1380 CNRS - CEA/DTA/DECM
   Ecole Polytechnique, F-91128 Palaiseau, France \\ and 
Istituto Nazionale per la Fisica della Materia, Dipartimento di Fisica
   dell'Universit\`a \\ di Roma Tor Vergata, 
   Via della Ricerca Scientifica, I--00173 Roma,
   Italy }
\author{R.W. Godby}
\address{  Department of Physics, University of York,
York YO1 5DD, United Kingdom}

\protect{\bigskip}
\bigskip
%\date{\today}
\draft
\maketitle
\bigskip
\begin{abstract}

We present a new method for the computation 
of self-energy corrections 
in large supercells. 
It eliminates
the explicit 
summation over unoccupied states, 
and uses an iterative
scheme based on an expansion of the Green's function
around a set of reference
energies. This improves the scaling of the computational time from
the fourth to the third power of the number of atoms for both the
inverse dielectric matrix and the self-energy, yielding
improved efficiency for 8 or more silicon atoms per unit cell.

\end{abstract}
%\twocolumn
\section*{}
\narrowtext
The quasiparticle (QP) band structure of a system of interacting electrons
(the single-particle-like
approximate eigenstates which describe the addition of an electron or a hole)
is obtained from solutions of a Schr\"odinger equation in which
exchange and correlation is described by
the electron self-energy $\Sigma$.
{\it Ab initio} calculations of QP energies
for real solids have been performed since the 1980s
[e.g. \cite {hanke,hyblou,godshsh}], generally using two
 approximations: (i) the self-energy is evaluated
within Hedin's $GW$ approximation \cite {hedin},
 where it is described as the
convolution of the one-particle Green's function $G$ and
 the screened Coulomb
interaction $W$, both of which are obtained from an initial
density-functional-theory (DFT) calculation using the 
local-density approximation (LDA); and
 (ii) the QP energies
are evaluated in first-order perturbation
theory, starting from the DFT-LDA eigenvalues and eigenstates.     
Band structures in excellent agreement with experiments
have been obtained in this way for many systems, but the 
applications are
at present restricted to relatively small basis sets and unit cells:
calculating the inverse dielectric matrix and the QP energies
is computationally demanding, and
scales essentially as $N_{at}^4$, 
the fourth power of the number of atoms in a supercell.  
The potential applications of {\it ab initio} electronic-structure calculations 
are therefore restricted in comparison with ground-state calculations.
\par
A recent real-space-imaginary-time approach
\cite{realsp} scales as $N_{at}^2$ for the
construction of the Green's function and as $N_{at}$ for the 
formation of the dielectric matrix and self-energy.  
However, the method is designed for calculations that require the 
whole self-energy $\Sigma ( {\bf r}, {\bf r'}, \omega )$ 
for all ${\bf r}$ and ${\bf r'}$, and, for feasible system sizes, 
is less efficient if only a few matrix
elements of $\Sigma$ are required.  A mixed-space approach \cite{mixedspace}
for the dielectric matrix scales as $N_{at}^3$, but does not address the 
construction of the self-energy.
In this paper we describe a new approach that yields efficient calculation of
QP energies for supercells of the order of 10-100 atoms.
\par
In the past, considerable  progress has been made in the 
calculation of the 
single-particle time-ordered
response function $\chi_0$,
\widetext
\begin{equation}
  {\chi^0}_{ {\bf G}  {\bf G'} }( {\bf q}, \omega) = 
  {2 \over \Omega }
  \sum_{n  n'  {\bf k}} (f_{n{\bf k}'} - f_{n'{\bf k}}) 
  { \langle n{\bf k}' \vert e^{- i ({\bf q} +
  {\bf G}) \cdot {\bf r}} \vert n'{\bf k} \rangle 
    \langle n'{\bf k} \vert e^{ i ({\bf q} +
  {\bf G'}) \cdot {\bf r}'} \vert n{\bf k}' \rangle
  \over
   \left( E_{n{\bf k}'} - E_{n'{\bf k}} + \omega + i  \eta \cdot
sgn
\left( E_{n'{\bf k}} - E_{n{\bf k}'}\right) \right)
   }
\end{equation}
\narrowtext
\noindent where $ {\bf k}' =  {\bf k} - {\bf q} $, $n$ and $n'$ run
over the bands, the $f_{n'{\bf k}}$ are occupation numbers,
 $\vert n{\bf k}' \rangle$ are the one-electron eigenfunctions
calculated, for instance, using the local-density approximation,
and $\Omega$ is the volume of the unit cell.
Based on the perturbation summation approach of Dalgarno and 
Lewis \cite{dalgarno},
Baroni and co-workers \cite{barqua,baroni} have designed a Green's-function approach 
which avoids
the explicit summation over unoccupied states in (1) for the case of 
static 
response; their method has recently been generalized  
to the dynamical
$\chi_0$ \cite {quong}. 
 It is hence a natural idea to extend
those advantages to self-energy calculations,
in which $\chi_0$ is the main ingredient for the determination of   
the inverse dielectric matrix
$\epsilon_{{\bf G},{\bf G}'}^{-1}({\bf q},\omega )$, and a 
second sum over
empty states, arising from the Green's function $G$, appears in the 
expression for the matrix elements
of $\Sigma$.
 At first sight the matrix elements of $\Sigma$ 
are formally similar to $\chi_0$, especially in a plasmon-pole approximation 
such as that of Ref. \cite {godneed}: 
essentially, (1) has to be 
modified by
some $({\bf G},{\bf G'})$-dependent prefactors, and $\omega $ 
substituted by the plasmon-pole parameters 
$~{\tilde\omega}_{{\bf G}{\bf G'}}$ (see later).
However, two main obstacles hinder the extension of the approach
to self-energy calculations.
First, when all matrix 
elements $({{\bf G},{\bf G'}})$ of $\chi_0$ are needed, as is 
the case in QP calculations,
 the straightforward application of the method  
proposed in Ref. \cite {quong} to $\chi_0$  still yields  an
 $N_{at}^4$
scaling:  this is because
the method requires a matrix inversion, which scales as  $N_{at}^3$, for each
of the $N_v$ energy denominators appearing in (1), where $N_v$, the number of valence
states, is obviously proportional to $N_{at}$. 
In the case of $\langle \Sigma \rangle$ the situation is even
worse, since the number of different energy denominators
is proportional to N$_v$ times the number of different plasmon
pole frequencies          
${\tilde\omega}_{{\bf G}{\bf G'}}$, one for each (G,G') pair. 
The scaling would be in this case 
$N_{at}^6$!

Here we propose
an extension of the Green's-function 
technique to
the calculation of self-energy corrections which maintains its 
advantages,
and moreover has the improved scaling of $N_{at}^3$ 
(ignoring log $N_{at}$ contributions)
for both $\epsilon_{{\bf G},{\bf G'}}^{-1}({\bf q},\omega )$ 
and $\langle \Sigma \rangle $, together with a favourable
prefactor. 
Our approach
is based on Taylor expansions of the Green's functions, which 
are shown to
converge rapidly, keeping the same numerical stability and 
controllability as
the commonly used empty-states summation method.
We illustrate the 
performance of our method for the example of successively 
large supercells
of bulk silicon,
showing that quasiparticle calculations in the 
framework
of the standard $GW$ approach \cite {hyblou,godshsh,hedin,godneed}
 for systems such as
point defects or amorphous silicon are made feasible.  
\par
We start from the Green's-function idea of Ref. \cite 
{barqua,quong} for the
calculation of $\chi_0$ in (1), where the solution of the linear 
system
\begin{equation}
      \left( -H + E_{v{\bf k}'} \pm \omega +i \eta \right)
  \vert \Psi^{\pm}_{v,{\bf k}', {\bf q}, {\bf G}', \omega}
    \rangle =
  e^{i ({\bf q} + {\bf G'}) \cdot {\bf r}} \vert v{\bf k}' \rangle 
\end{equation}
\noindent for the ``polarization state''
$\vert \Psi^{\pm}_{v,{\bf k}', {\bf q}, {\bf G}', \omega}\rangle $ 
allows to rewrite  $\chi_0$ as

\begin{equation} 
  \chi^0_{ {\bf G} \, {\bf G'}}({\bf q}, \omega) =
   {2 \over \Omega }
 \sum_{{\bf k} \, v }{ \langle v,{\bf k}' \vert
 e^{- i ({\bf q} + {\bf G})
 \cdot {\bf r}}
   (\vert \Psi^{+} \rangle  + \vert \Psi^{-} \rangle )}
 \end{equation}

\noindent As above, $  \vert v{\bf k} \rangle  $ is the LDA Bloch function
 $  e^{ i {\bf k} \cdot {\bf r}}
 u_{v,{\bf k} } ({\bf r})$,   $ {\bf k}' =  {\bf k} -  {\bf q} $, and $H$ is
 the LDA Hamiltonian.

  When  $\omega$ tends to zero,
 $ ( -H + E_{v,{\bf k}} \pm \omega +i \eta )^{-1}$
 diverges, as do $\vert \Psi^{+} \rangle $ and  $\vert \Psi^{-} \rangle $.  
 However,
the sum (3) remains finite, since $\langle \Psi^{-}_{v',{\bf k}, -{\bf q}, 
-{\bf G}', \omega}\vert v{\bf k}'\rangle = - \langle v'{\bf k}\vert 
\Psi^{+}_{v,{\bf k}', {\bf q},
{\bf G}', \omega} \rangle $.
   For numerical stability, it is then better to   
project out the valence states from $H$ since the beginning,
 in a way
  similar to that of ref.\cite{barqua}:
 we change the operator $H$ appearing
 on the left side of Eq. (2) into $\tilde H = H P$ where $P = 1 -
 \sum_{v} \vert v \rangle \langle v \vert $,
 and we modify the right--hand side of Eq. (3) inserting the
 projector $P$ to the left of both $\Psi^{+}$ and $\Psi^{-}$.

We write (2) in reciprocal space as 
\widetext
\begin{equation}
  \sum_{{\bf G}''}   \left( -\tilde H_{{\bf k} + {\bf G}, {\bf k} +
  {\bf G}''} + ( E_{v,{\bf k}'} \pm \omega +i \eta)
  \delta_{{\bf G}, {\bf G}''} \right)
  f^{\pm}_{{\bf G}''} (v, {\bf k}',{\bf q},{\bf G'},\omega) =
  u_{{\bf G} - {\bf G}' }( {v,{\bf k}') }  \eqnum{2b} 
\end{equation}
\narrowtext

\noindent where $f^{\pm}({\bf r}) = e^{- i {\bf k}  \cdot {\bf r}}
 \Psi^{\pm}_{v,{\bf k}',{\bf q},{\bf G}',\omega} ({\bf r})$ is a
 periodic function in ${\bf r}$.
 An LU
  decomposition (or inversion) of $ (-\tilde H + E_v \pm \omega + i \eta )$ 
scales as $N^3$ with the number
of plane waves, $N$,  for every valence energy $E_v$,
 yielding the $N_{at}^4$ scaling of the 
``na\"ive'' implementation.
 Then, the solution for each right-hand side requires 
 $N^2$ operations for every $G'$ and every $  u_{v,{\bf k}'} $
(or $N^2logN$ operations for every $  u_{v,{\bf k}'} $ and the whole set
of $G'$, when fast Fourier transforms (FFTs) are used).

  Equation (3) becomes 
 \widetext
 \begin{equation}
   \chi^0_{ {\bf G} \, {\bf G'}}({\bf q}, \omega) =
 {2\over \Omega} ~\sum_{v {\bf k}'} ~
  \int d {\bf r} ~~ u^*_{v,{\bf k}'}({\bf r}) e^{- i {\bf G}  \cdot {\bf r}} 
 (f^{+}_{\bf r}(v,{\bf k}',{\bf q},{\bf G}',\omega ) + 
 f^{-}_{\bf r}(v,{\bf k}',{\bf q},{\bf G}',\omega )), \eqnum{3b} 
 \end{equation}
 \narrowtext\noindent
which is again computed efficiently using FFTs.
Before coming to the possible
 improvements in the calculation of $\chi_0$,
it is useful to look directly at the self-energy matrix elements.
The main numerical effort lies in the evaluation of
 the correlation contribution,
which is
\widetext
\begin{eqnarray}
 \langle n {\bf k}\vert \Sigma_c (\omega )\vert n {\bf k} \rangle &=& 
{2  \pi \over {N \Omega}}
 \sum_{c'{\bf k}'} \sum_{{\bf G} {\bf G}'}
 {\Omega^2_{{\bf G}{\bf G}'} \over { |
 {\bf q} + {\bf G} |  |
  {\bf q} + {\bf G}' |} } { \langle n {\bf k} \vert
 e^{- i ({\bf q} + {\bf G}) \cdot {\bf r}} | c'{\bf k}' \rangle
  \langle c'{\bf k}' | e^{ i ({\bf q} + {\bf G'}) \cdot {\bf r}'}
 \vert n{\bf k} \rangle \over 
{ \tilde\omega_{{\bf G}{\bf G}'}({\bf q})
 \left( \omega - \tilde\omega_{{\bf G}{\bf G}'}({\bf q}) 
  - E_{c'{\bf k}'} \right) }} \nonumber \\ 
&+&  {{2  \pi \over {N \Omega}}
 \sum_{v'{\bf k}'} \sum_{{\bf G} {\bf G}'}
 {\Omega^2_{{\bf G}{\bf G}'} \over { |
 {\bf q} + {\bf G} |  |  {\bf q} +
 {\bf G}' |} }{ \langle n {\bf k} \vert
 e^{- i ({\bf q} + {\bf G}) \cdot {\bf r}} | v'{\bf k}' \rangle
  \langle v'{\bf k}' | e^{ i ({\bf q} + {\bf G'}) \cdot {\bf r}'}
 \vert n{\bf k} \rangle \over 
{ \tilde\omega_{{\bf G}{\bf G}'}({\bf q})
 \left( \omega + \tilde\omega_{{\bf G}{\bf G}'}({\bf q})
  - E_{v'{\bf k}'} \right) }}} 
\end{eqnarray}
\narrowtext
\noindent where $ {\bf k}' =  {\bf k} - {\bf q} $, 
  $\Omega_{{\bf G}{\bf G}'} $ and
 $ \tilde\omega_{{\bf G}{\bf G}'}$ are
  the plasmon-pole parameters 
determined by the energy dependence of
$\epsilon_{{\bf G},{\bf G}'}^{-1}({\bf q},\omega )$, 
and $c'$ ($v'$) denote
sums over unoccupied (occupied) states. 
Because of the sum over unoccupied states, the first of the two terms is the 
computationally demanding one.
Contrarily to eq.(1), here there is a different energy denominator
for each plasmon--pole frequency, $\tilde\omega_{{\bf G}{\bf G}'}$.
As pointed out above, performing a new LU decomposition (or inversion)
for each individual ($G$,$G$') pair would lead to an $N_{at}^6$ scaling,
 making 
the approach disadvantageous even with respect to the 
traditional method. 
However,it is known that  
 $\Sigma (\omega)$ is a smooth function of $\omega$, which implies
that it might also be a smooth function of $\omega - 
\tilde\omega_{{\bf G}{\bf G}'}$.
 We observe that many of the $N^2$ different
 $ \tilde\omega_{{\bf G}{\bf G}'}$ have similar values, and that $\omega$
is typically taken at the DFT eigenvalue of state $\vert nk\rangle $;
it is therefore 
possible to evaluate the contributions from different 
 $ \tilde\omega_{{\bf G}{\bf G}'}$ by means of a Taylor expansion of
 $ (\omega - \tilde\omega_{{\bf G}{\bf G}'} -\tilde H )^{-1}$ 
around $ \omega - \tilde\omega_{{\bf G}{\bf G}'}$ = $ 
E_{n,0} -\tilde\omega_0$, for one
or more expansion points $E_{n,0}$ and $ \tilde\omega_0$ in the range of
interest.
Since the width of the energy interval into which the 
$E_{n}$ of interest and the plasmon pole parameters
 $ \tilde\omega_{{\bf G}{\bf G}'}$
are scattered does not grow with the system size, the number of
expansion points needed -- and hence the number of matrix inversions -- no
longer depends on N$_{at}$.

Expression (4) also contains possibly
 divergent contributions, which could hinder the 
application of the Green's-function approach to the first term:
when  E$_{c'{\bf k}'}$ is an unoccupied state,
divergences arise for $ \omega - E_{c'{\bf k}'} = \tilde \omega $,
hence when $ \omega $ is at an energy at least
$\tilde\omega_{{\bf G}{\bf G}'}$ {\it above} the
 lowest conduction state.
In order to avoid such divergences, 
 we divide the unoccupied states into two groups: the `low--lying'
states with energies from the Fermi energy to somewhat beyond the 
highest quasiparticle energy of interest, and the
states with higher energies.  Then we apply the Green's-function trick to
the latter states only, now including the low--lying states
in the projector $P$,
making the solution well defined  for all the 
energies in the range of interest.

The Taylor coefficients, i.e. the energy derivatives of
$ \tilde G(\tilde\omega) =  (- H P + E_{0,n} - \tilde\omega) ^{-1}$, 
are essentially 
powers of  $\tilde G(\tilde\omega_0)$. Then, their computation
only requires 
as many matrix multiplications as the order of the expansion 
(typically 1 or 2).
The contributions of the different orders can be evaluated
separately, in a CPU time  
proportional to $N^3_{at}$ \cite{nota1}.
The derivative $\langle \partial
 \Sigma /\partial \omega \rangle$, used in calculating $\langle n\vert
\Sigma (\omega)\vert n \rangle $ as $\langle n\vert \Sigma
(E^{DFT}_n)\vert n\rangle + \langle \partial \Sigma /\partial
 \omega \rangle \cdot (\omega-E^{DFT}_n)$ (see Ref. \cite{godneed}),
is also determined by the second power of
$\tilde G$, and hence does not require any further effort.
The only term whose evaluation is in principle still
proportional to $N^4_{at}$ is the summation over the occupied and
lowest unoccupied states, which is performed explicitly.
However, the actual
number of operations involved  
is negligible, since the number of these low--lying  
states is small.

The expansion approach can be used in the same way to improve the 
calculation of $\chi_0 (\omega) $. Here there are
in principle as many different energies in the denominator as
valence states. Using the expansion technique, one only has to
 compute
$(-\tilde H + E_{v0} + \omega)^{-i} $ for a number of
 expansion points $E_{v0}$ which
is proportional to the width of the valence band (independent of system size). 
In the case of $\chi_0$ for silicon,
inclusion of orders $(i-1)$ from 0 to 3 is
sufficient to achieve an accuracy of about 50 meV 
in $ \langle \Sigma \rangle $, using 3 expansion points.

To demonstrate that the method is 
well suited to {\it ab initio} $GW$ calculations for large 
unit cells
with many electrons, we have computed
the  $GW$-corrected electronic 
structure for
different supercells of bulk silicon, 
using the $\Gamma$ point only. In Table I, we illustrate 
for the 2-atom cell the good convergence of 
single elements of $\epsilon^{-1}$ with the order of the 
Taylor expansion. ``Exact'' results are obtained by using the traditional
method, with 181 plane waves (which corresponds to the 12.5 Ry energy cutoff
used in the DFT calculation) and the whole 181 states. Table II shows
the results for the matrix elements of $\Sigma$ as a function of the
order of its Taylor expansions (using a 3rd order expansion for
$\epsilon^{-1}$ throughout). We compare with the exact results and with the approximate
results which are obtained with 
the traditional approach, using 169 plane waves and 112 
states.
The `traditional' results have an accuracy of 50 meV, and, with our method, inclusion 
of orders 0, 1 and 2 are seen to be  
sufficient to achieve the same accuracy.

The computational time as a function of the supercell size is
illustrated in Fig. 1. To allow a fair comparison,
both calculations
have been performed with the parameters leading to an accuracy of
about 50 meV. 
The two approaches are already equivalent for an 
8-atom silicon supercell.  With 54 atoms the new method
gains a factor of five in computing $\langle\Sigma\rangle $;
here the CPU time required to
compute the dielectric function,
 plus $\Sigma$ and its energy derivative
for all 66 states lying within 0.1 Hartrees of the Fermi
 energy is 42 hours on a Cray C98 computer, 22 hours of which is for $\chi_0$. 
This means that
quasiparticle calculations in the framework
of the standard $GW$ approximation for systems such as
point defects or amorphous silicon are feasible 
with a reasonable computational effort.
We have already successfully applied the present approach
 to the calculation
of  $\chi_0$ and 
$ \langle \Sigma \rangle $ for sodium clusters in a large
supercell\cite{na4}.

A possible improvement, which will reduce the
 computational effort further,
is to exploit the fact that in our scheme, by
analogy with what is pointed out in Ref. \cite{realsp} and
\cite{mixedspace}, we compute quantities
which are short-ranged in $({\bf r} - {\bf r}')$.  
Introducing a real-space cutoff in $ ( -H P + \omega - \tilde \omega )^{-1} $
will reduce the computational expense significantly (at 
present, without the cutoff, FFTs account for
more than half of the total CPU time.)

 In summary, 
 we have introduced a scheme which extends towards larger
 sizes and complexity the set of physical systems 
 for which self-energy-corrected electronic structure can be computed,
 within Hedin's $GW$ scheme and the plasmon-pole approximation.
 By avoiding the need for an explicit summation
 over conduction states, 
 and introducing an iterative
 scheme to 
 describe the energy dependence of the
 Green's function by expanding around a few reference energies,
 the method reduces the scaling of the computational time from
 $N^4_{at}$ to $N^3_{at}$. Systems with a number of atoms $N_{at}$ of the order 
 of 50, which would require a prohibitive computational effort 
 within traditional $GW$ schemes, thereby become accessible.
 The present method appears to be a promising tool for the study 
 of complex structures such as
 clusters, reconstructed surfaces, and 
 point defects in semiconductors, since they often fall within
 this class of systems. 

This work was supported in part by the European Community programme 
``Human Capital and Mobility'' through Contract No. ERB CHRX CT930337. 
Computer time on the Cray C98 was granted by IDRIS (Project No. CP9/950544).

\newpage

\widetext
\begin{table}
\caption{Typical convergence of (${\bf G},{\bf G'}$) elements of $\epsilon^{-1}$
with the order of the Taylor expansion (see text).}
\begin{tabular}{ccccccccccc}
{\bf G} & {\bf G}' &     order 0    &  1  &    2  &    3  &    4  &
    5 &    6 &     7  &   exact \\
\tableline
($000$) & ($000$)  &       0.0347 &   0.0121 & 0.0082 & 0.0069 &0.0064 &
  0.0061  &    0.0060 &    0.0060 &    0.0059 \\
(${\bar 1}1{\bar 1}$) & (${\bar 1}1{\bar 1}$)   &     0.6373  &  0.6197 &
0.6138&  0.6121& 0.6114 &  0.6112   &   0.6111  &   0.6110   &  0.6110 \\
($00{\bar 2}$) & ($00{\bar 2}$)  &      0.6336  &  0.5998 & 0.5879 &
 0.5833 &0.5812 &  0.5802 &    0.5797  &  0.5794  &    0.5792 \\
($111$) & ($00{\bar 2}$)  & -0.0215\tablenote{times $1-i$ } &
-0.0231\tablenotemark[1] &-0.0240\tablenotemark[1]  &
  -0.0244\tablenotemark[1]  & -0.0246\tablenotemark[1] &
  -0.0247\tablenotemark[1]   &   -0.0247\tablenotemark[1]  & 
  -0.0248\tablenotemark[1]  &   -0.0248\tablenotemark[1] \\
($3{\bar 1}1$) & (${\bar 3}1{\bar 1}$) &  0.0131\tablenote{times $i$ }   & 
 0.0167\tablenotemark[2]  &  0.0183\tablenotemark[2]   & 
 0.0190\tablenotemark[2]  &  0.0193\tablenotemark[2] &
 0.0195\tablenotemark[2]   &  
0.0195\tablenotemark[2]   &  0.0196\tablenotemark[2]  & 
  0.0196\tablenotemark[2] \\
\end{tabular}
\label{converg_chi}
\end{table}
\narrowtext
\begin{table}
\caption{ Quasiparticle corrections to the LDA eigenvalues for
 bulk silicon, calculated with the present
 method for different 
%numbers of iterations
orders of the Taylor expansion for $\Sigma$
 (columns 1, 2 and 3), and with the 
traditional method
 summing over  {\it all} the conduction states (for the chosen plane-wave basis set)
 (column 4), and summing over 
about 2/3 of the conduction states (column 5). 
Values in eV.         } 

\begin{tabular}{cccccc}
& 1$^{st}$ ord. & 2$^{nd}$ ord.  & 3$^{rd}$ ord.  & ``Exact'' & Trad.\\
\tableline
$\Gamma^{'}_{25v}$  &  0.080 &  0.030  & 0.020  & -0.013 & 0.035 \\
$\Gamma_{15c}$  &  0.576 &  0.515  & 0.498  & 0.485 & 0.530 \\
$\Gamma^{'}_{2c}$  &  0.692 &  0.605  & 0.575  & 0.553 & 0.582 \\
\end{tabular}
\label{converg_sig}
\end{table}
\narrowtext
\newpage
\narrowtext
\begin{figure} 
\caption { CPU time required for the calculation
 of the $GW$ corrections to the LDA electronic structure,  
 as a function of the number of atoms in the supercell. Filled diamonds are
 for the traditional method, hollow squares for the present scheme. 
 In the inset, a log-log plot of the same variables
 shows the cross-over at the 8 atom supercell. \label{cputimefig}}
\label{fig:cputime}
\end{figure}


\bigskip

\newpage
 
\begin{references}

\bibitem{hanke} G. Strinati, H. J. Mattausch, and W. Hanke, Phys Rev. B {\bf
25}, 2867 (1982).
\bibitem{hyblou}M.S. Hybertsen and S.G. Louie,
 Phys. Rev. Lett. {\bf 55},
1418 (1985), Phys. Rev. {\bf B 34}, 5390 (1986).
\bibitem{godshsh}  R.W. Godby, M. Schl\"{u}ter, and L.J. Sham,
 Phys. Rev. Lett. {\bf 56}, 2415 (1986); Phys Rev. B {\bf37}, 10159
(1988).
\bibitem{hedin} L. Hedin, Phys. Rev. {\bf 139}, 796 (1965).
\bibitem{realsp}  H.N. Rojas, R.W. Godby, and R.J. Needs, Phys. Rev.
 Lett. {\bf 74} , 1827 (1995).  
\bibitem{mixedspace}  X. Blase, A. Rubio, S.G. Louie, and M. Cohen,
Phys. Rev.  B. {\bf 52}, R2225, (1995).
\bibitem{dalgarno}A. Dalgarno and J.T. Lewis, Proc. R. Soc.
 London {\bf A 233}, 70 (1955).
\bibitem{barqua} S. Baroni and A. Quattropani, 
Il Nuovo Cimento {\bf 5D}, 89 (1985).
\bibitem{baroni}  S. Baroni, P. Giannozzi, and A. Testa,
 Phys. Rev. Lett.
{\bf 58}, 1861 (1987).
\bibitem{quong} A.A. Quong and A.G. Eguiluz, Phys. Rev. Lett.  {\bf
70}, 3955 (1993).
\bibitem{godneed} R.W. Godby and R.J. Needs,
 Phys. Rev. Lett. {\bf 62},
1169 (1989).
\bibitem{nota1} Matrix multiplications are performed using the
 Strassen
  algorithm, which scales as $N^{2.8}$  rather than $ N^3 $.
\bibitem{na4} G. Onida, L. Reining, R.W. Godby, R. Del Sole,
 and W. Andreoni,  Phys. Rev. Lett. {\bf 75}, 818 (1995).



\end{references}
\end {document}
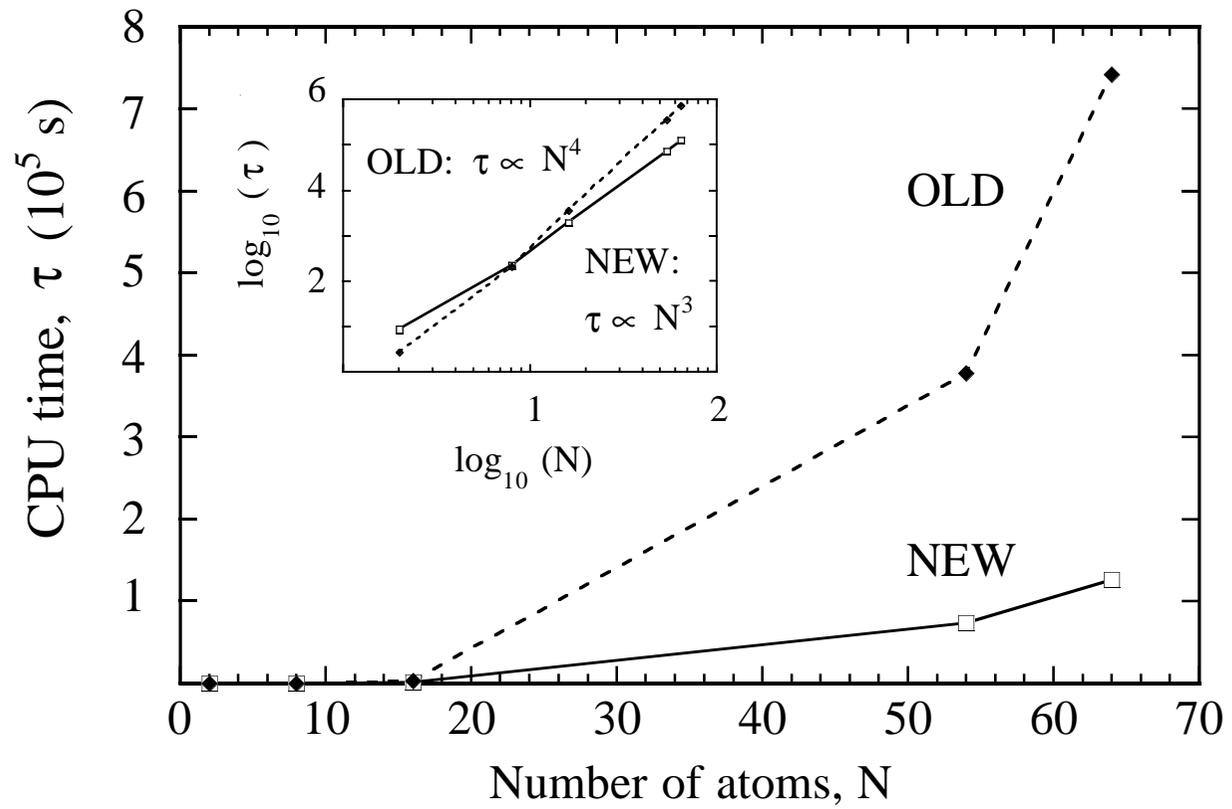
